\documentclass[twocolumn,superscriptaddress,prl]{revtex4}
\usepackage{amsmath}

\usepackage{graphicx}% Include figure files
\usepackage{dcolumn}% Align table columns on decimal point
\usepackage{bm}% bold math
\usepackage[version=4]{mhchem}
\usepackage{verbatim}

\usepackage{xcolor} %color stuff

\newcommand{\md}{\mathrm{d}}
\newcommand{\st}{\boldsymbol{\phi}} %state variable
\newcommand{\St}{\boldsymbol{\Phi}} %state variable capital
\newcommand{\org}{s_{\rm{-}}} %outcome variable
\newcommand{\Org}{S_{\rm{-}}} %outcome variable capital
\newcommand{\out}{s_{\rm{+}}} %origin variable
\newcommand{\Out}{S_{\rm{+}}} %origin variable capital
\newcommand{\meso}{m} %mesostate label
\newcommand{\Meso}{\mathcal{M}} %set of mesostates
\newcommand{\Nmeso}{N_{\rm meso}} %Number of mesostates
\newcommand{\subens}{\boldsymbol{s}} %subensemble variable
\newcommand{\Subens}{\boldsymbol{S}} %subensemble variable capital
\newcommand{\RC}{r} %reaction coordinate label

\begin{document}
	
\preprint{APS/123-QED}
	
\title{Information thermodynamics of transition paths between multiple mesostates}
	
\author{Miranda D.\ Louwerse}
\altaffiliation{Current address: AbCellera Biologics Inc., Vancouver, British Columbia, Canada}
\affiliation{%
Department of Physics, Simon Fraser University, Burnaby, British Columbia, Canada V5A1S6
}%
\author{David A.\ Sivak}%
\email{dsivak@sfu.ca}
\affiliation{%
Department of Physics, Simon Fraser University, Burnaby, British Columbia, Canada V5A1S6
}%

\date{\today}
	
\begin{abstract}
A central concern across the natural sciences is a quantitative understanding of the mechanism governing rare transitions between two metastable states. Recent research has uncovered a fundamental equality between the time-reversal asymmetry of the ensemble of such transition paths and the informativeness of system dynamics about the reactivity of a given trajectory, immediately leading to quantitative criteria for judging the importance of distinct system coordinates for the transition. Here we generalize this framework to multiple mesostates. We find that the main system-wide and coordinate-specific results generalize intuitively, while the combinatorial diversity of pairwise transitions raises new questions and points to new concepts. This work increases the previous framework's generality and applicability and forges connections to enhanced-sampling and coarse-grained dynamical approaches such as milestoning and Markov-state models.
\end{abstract} 
	
\maketitle

\section{Introduction}
Biomolecular dynamics are studied and understood on several length scales, from atomic details to more coarse-grained aspects~\cite{Prinz2011}. The mechanisms by which reactions occur are typically couched in terms of a global picture of the dynamics, including distinguishable metastable states, their relative stability, and transitions between them. At the same time, microscopic details that affect the stability of mesostates or the transitions between them are of interest, especially in applications where tuning molecular components affects the function of the molecule~\cite{Peters2016}. It is therefore desirable to build models of dynamics that capture both global and local features of the system.

Much of the theory of reaction dynamics is developed in the context of a reaction between two endpoints, which from a coarse-grained perspective is essentially two-state, parameterized by forward and reverse rate constants, which can be derived from microscopic dynamics using transition-path theory~\cite{Metzner2009,E2006}. In previous work~\cite{Louwerse2022}, we used tools from stochastic thermodynamics to adapt the transition-path theory description to propose an information-theoretic approach, in which an irreversible entropy-production rate quantifies the informativeness of motion in relevance of microscopic degrees of freedom to describing the mesoscopic reaction dynamics, i.e., to determining the rate constant. Through this, we showed that in this information-theoretic sense the committor function, a central quantity in transition-path theory, is a sufficient reaction coordinate (capturing all available system information about trajectory reactivity) and therefore the ``true" reaction coordinate~\cite{Peters2016}. 

Here, we extend these ideas to reactions between multiple mesostates. These types of reaction networks are commonly seen in biomolecular systems and are frequently modeled using coarse-grained Markov-state models~\cite{Chodera2014,Pande2010,Prinz2011,Schutte2011,Buchete2008}. We define an information-theoretic description of a multi-mesostate reaction network and relate the informativeness of system dynamics to the entropy-production rate for reactive dynamics, in close analogy to the relation derived for two endpoints in our previous work~\cite{Louwerse2022}. However, the multi-mesostate model has some nuances: e.g., the committor (defining the probability of next reaching a particular mesostate given an initial microstate) becomes a vector~\cite{Schutte2011}; we show that this committor is a sufficient statistic for relating the reaction network to the microscopic dynamics. We also derive an entropy-production rate for transition-path ensembles between each pair of mesostates and derive a local reaction coordinate for each pairwise reaction that fully captures this entropy-production rate. We discuss how these information-theoretic quantities could be used to build informative global models of the reaction network while at the same time highlighting system features important to each local reaction. Finally, we show that the equating of entropy production and information generation also holds for subensembles of trajectories that leave a given mesostate~\cite{Berezhkovskii2019} (without specifying the destination mesostate). With this multi-mesostate framework, the main results from our previous work generalize simply and intuitively, increasing their applicability.

\section{Categorical outcome/origin variables and trajectory subensembles}
We first generalize our theoretical framework to accommodate more than two mesostates of interest.

\subsection{System dynamics}

A system with discrete microstates $\st$ in configuration space $\St$ evolves under Master-equation dynamics 
\begin{equation}
\md_t p(\st) = \sum_{\st'} T_{\st \st'} p(\st')
\end{equation}
with probability $p(\st)$ of microstate $\st$ and $\st' \to \st$ transition rate $T_{\st \st'}$. The system has reached the equilibrium Boltzmann distribution 
\begin{equation} \label{eq:Boltzmann_distr}
\pi(\st) = e^{\beta F-\beta E(\st)} \ ,
\end{equation}
for energy $E(\st)$ of microstate $\st$ and equilibrium free energy $F \equiv - \ln \sum_{\st} e^{-\beta E(\st)}$. At equilibrium the probability of microstate $\st$ does not change, 
\begin{equation}
0=\md_t \pi(\st) = \sum_{\st'} T_{\st \st'} \pi(\st') \>,
\end{equation}
and detailed balance
\begin{equation} \label{eq:microstate_detailed_balance}
T_{\st \st'} \pi(\st') = T_{\st' \st} \pi(\st)
\end{equation}
holds for each individual $\st'\to\st$ transition. 

Define a set of $\Nmeso$ mesostates $\Meso = \{ \meso_i \}$, each of which is a distinct subset of microstates (e.g., in the two-mesostate transition-path theory~\cite{E2006}, $\Nmeso=2$, $\meso_1=A$, and $\meso_2=B$). (To simplify the analysis, we assume there are no direct transitions between distinct mesostates.) These mesostates could be chosen as the basins of metastable states~\cite{Chodera2014}, at transient ``milestones'' during a transition~\cite{Elber2004}, or in other ways. For each time in a long equilibrium ``supertrajectory'', we define the trajectory outcome $\Out$ and origin $\Org$ as respectively the next and the most recent mesostate visited by the system, which are categorical random variables with values drawn from the list of mesostates: $\out,\org \in \{ \meso_i \}$. The joint distribution of microstate $\st$, trajectory origin $\org$, and trajectory outcome $\out$ can be decomposed as
\begin{equation} \label{eq:joint_distr_st_org_out}
p(\st,\org,\out) = \pi(\st) p(\org|\st) p(\out|\st) \>.
\end{equation}
The right-hand-side decomposition relies on the conditional independence of trajectory outcome and origin given a microstate $\st$, which follows from the Markovianity of the underlying dynamics~\cite{E2006,Metzner2009}.

By marginalizing over all microstates $\st$ the joint distribution, the supertrajectory is partitioned into $\Nmeso^2$ trajectory subensembles $\subens=(\org,\out)$ with marginal probability distribution
\begin{equation} \label{eq:marg_distr_org_out}
p(\org,\out) = \sum_{\st} \pi(\st) p(\org|\st) p(\out|\st) \>.
\end{equation}
There are $\Nmeso$ `stationary' subensembles (where the system leaves and returns to the same mesostate) and $\Nmeso^2-\Nmeso$ `reactive' subensembles (where the system leaves one mesostate and next reaches another). The forward and reverse transition-path ensembles between any pair of mesostates are time-reversed counterparts with equal-magnitude flux of reactive trajectories at any given point in configuration space; there are therefore ${\Nmeso \choose 2} = \Nmeso(\Nmeso-1)/2$ ``distinct'' reactive subensemble pairs.

\subsection{Forward and reverse multi-mesostate committor vectors}
The conditional probability of a trajectory outcome (origin) given a microstate $\st$, $p(\Out = \meso_i|\st)$ [$p(\Org = \meso_i|\st)$] (appearing in Eq.~\eqref{eq:joint_distr_st_org_out}), is a multi-mesostate generalization of the forward (backward) committor function that appears in the two-mesostate transition-path theory: it gives the probability that a system initiated at $\st$ will next reach (last visited) mesostate $\meso_i$~\cite{Schutte2011}. The forward (backward) committor $p(\out|\st)$ [$p(\org|\st)$] for multiple mesostates is a vector function of dimension $\Nmeso$, with each element giving the probability the system will next visit (last visited) mesostate $\meso_i$~\cite{Schutte2011}, with normalization
\begin{subequations} \label{eq:committor_sum}
\begin{align}
\sum_{i=1}^{\Nmeso} p(\Out=\meso_i|\st) &= 1 \\
\sum_{i=1}^{\Nmeso} p(\Org=\meso_i|\st) &= 1 \>,
\end{align}
\end{subequations}
for each microstate $\st$. The traditional committor for a two-mesostate model can also be considered a vector function with two elements, but the constraints in Eq.~\eqref{eq:committor_sum} imply that both elements are determined by a scalar (because they sum to 1), thus the two-mesostate committor is conventionally described as a scalar function.

For each microstate $\st$, each element of the forward committor vector $p(\Out=\meso_i|\st)$ satisfies~\cite{Metzner2009,Schutte2011}: 
\begin{equation} \label{eq:multinomial_fwd_committor_recursion}
    0 = \sum_{\st'} p(\Out=\meso_i|\st') T_{\st' \st} \>,
\end{equation}
and the reverse committor vector $p(\Org=\meso_i|\st)$ satisfies
\begin{subequations}  \label{eq:multinomial_rev_committor_recursion}
\begin{align}
0 &= \sum_{\st'} p(\Org=\meso_i|\st') T_{\st \st'} \frac{\pi(\st')}{\pi(\st)} \\ 
&= \sum_{\st'} p(\Org=\meso_i|\st') T_{\st' \st} \label{eq:q_rev_eqm_assumption}
\end{align}
\end{subequations}
with boundary conditions $p(\Out=\meso_i|\st)=p(\Org=\meso_i|\st)=1$ for $\st \in \meso_i$ and $p(\Out=\meso_i|\st)=p(\Org=\meso_i|\st)=0$ for $\st \in \Meso \setminus \meso_i$. Equation~\eqref{eq:q_rev_eqm_assumption} uses the detailed-balance relation~\eqref{eq:microstate_detailed_balance}. For our time-reversible dynamics, the system of equations for $p(\Org=\meso_i|\st)$ and $p(\Out=\meso_i|\st)$ are identical, which implies $p(\Org=\meso_i|\st)=p(\Out=\meso_i|\st)$.

\section{Information theory}

The coarse-grained mesostate model just described retains some information about the system's microstate. This is quantified by the mutual information between system microstate and trajectory type, 
\begin{subequations}
\begin{align} \label{eq:mutual_information}
I[\St;\Subens] &= \sum_{\st, \subens} p(\st,\subens) \ln \frac{p(\st,\subens)}{\pi(\st)p(\subens)} \\
&= H(\Subens)-H(\Subens|\St) \label{eq:MI_reduce_sub}\\
&= H(\St)-H(\St|\Subens) \label{eq:MI_reduce_st} \>.
\end{align}
\end{subequations}
This is the reduction of uncertainty about the current trajectory type upon observation of the system microstate, or conversely the reduction of uncertainty about the microstate from knowledge of the trajectory type. For example, simulation methods for importance sampling of transition paths between two mesostates generate a set of trajectories~\cite{Dellago1998} or sample the flux of trajectories~\cite{Elber2004,vanErp2003,Allen2005} of a given trajectory type. This generates microstate probability distributions conditioned on each trajectory type (e.g., $A \to A$, $A \to B$, $B \to B$, $B \to A$) whose average entropy over all subensembles is less than the system entropy in the full equilibrium ensemble; the reduction in entropy is the mutual information between the system microstate and trajectory type (Eq.~\eqref{eq:MI_reduce_st}). By resolving one of these variables, either the current trajectory type or the system microstate, information is gained about the other. 

A major focus of chemical-physics research is the identification of a small set of coordinates that---despite vastly reduced dimensionality---can both distinguish between mesostates of interest and describe the progress of important reactions in the system~\cite{Peters2016}. For two-mesostate systems, it is well-recognized that this amounts to finding a small set of collective variables (functions of microstate) that can parameterize the committor function. Posed information theoretically, the mutual information between the reduced space of collective variables and trajectory type measures the sufficiency of the chosen collective variables to quantify the statistical relationship between the microstate and trajectory type (which determines the reactivity or stationarity of the current trajectory segment). From the data-processing inequality~\cite{Cover2006}, the mutual information between collective variables and trajectory type is less than or equal to the mutual information between microstate and trajectory type, and therefore information about the reactivity of the system is lost upon insufficient coarse-graining. An optimal set of collective variables preserves the mutual information upon coarse-graining. The forward committor vector has $\Nmeso$ elements, with a constraint that all elements must sum to unity; due to time-reversal symmetry, no further collective variables are needed to determine the reverse committor vector. An $\Nmeso$-mesostate model therefore requires at minimum $\Nmeso-1$ degrees of freedom to sufficiently quantify the information between underlying microstate and trajectory type. 

Let $\bm{q}=\{ q_i\}$ be an $\Nmeso$-dimensional vector with each entry the forward (or equivalently reverse) committor for mesostate $i$, constrained to values such that $\sum_i q_i = 1$. We show that the committor is a sufficient statistic for the mutual information between the microstate and trajectory type, as follows:
\begin{subequations}
\begin{align}
I[\St &;\Subens] = \sum_{\st, \subens} \pi(\st) p(\subens|\st) \ln \frac{p(\subens|\st)}{p(\subens)} \\
&= \int \md \bm{q} \sum_{\st, \subens} \pi(\st) p(\subens|\st) \delta[p(\out|\st)-\bm{q}] \ln \frac{p(\subens|\st)}{p(\subens)} \\
&= \int \md \bm{q} \sum_{i, j} q_i q_j \ln \frac{q_i q_j}{p(\Org=\meso_i,\Out=\meso_j)} \times \nonumber \\
&\quad\quad \sum_{\st} \pi(\st) \delta[p(\out|\st)-\bm{q}] \label{eq:group_q} \>,
\end{align}
\end{subequations}
where we used $p(\Org=\meso_i,\Out=\meso_j|\st)=q_i q_j$ and grouped together all microstates with the same committor-vector value. For marginal committor probability $\pi(\bm{q}) = \sum_{\st} \pi(\st) \delta[p(\out|\st)-\bm{q}]$, this further simplifies to
\begin{subequations}
\begin{align}
I[\St;\Subens] &= \int \md \bm{q} \sum_{i, j} \pi(\bm{q}) q_i q_j \ln \frac{\pi(\bm{q}) q_i q_j}{\pi(\bm{q}) p(\Org=\meso_i,\Out=\meso_j)} \\
&= I[\bm{Q};\Subens] \label{eq:MI_q_s} \>,
\end{align}
\end{subequations}
where we expressed the mutual information between microstate and trajectory type in terms of the sufficient statistic $\bm{Q}$, the random variable given by the committor of the microstate.

When the mesostates fully partition the configuration space, the committor vector of each microstate has unity for the mesostate to which the microstate belongs and zeroes for all other mesostates ($p(\subens=(\meso_i,\meso_i)|\st) = 1$ for $\st \in \meso_i$ and $0$ otherwise), so the mutual information between system microstate and trajectory type reduces to the trajectory-type entropy:
\begin{subequations}
\begin{align}
I[\St;\Subens] &= \sum_{\st} \sum_{\subens} \pi(\st) p(\subens|\st) \ln \frac{p(\subens|\st)}{p(\subens)} \\
&= -\sum_{i=1}^{\Nmeso} \left[ \sum_{\st \in \meso_i} \pi(\st) \right] \ln \left[ \sum_{\st \in \meso_i} \pi(\st) \right] \\
&= -\sum_{i=1}^{\Nmeso} \pi(\meso_i) \ln \pi(\meso_i) \\
&= H(\Subens) \>,
\end{align}
\end{subequations}
for Boltzmann probability $\pi(\meso_i)$ of the mesostate. In this case, observation of system microstate fully determines the current trajectory type, thus $H(\Subens|\St)=0$. If the number $\Nmeso$ of mesostates approaches the number of system microstates, knowing the microstate fully determines the trajectory type and vice versa, so the mutual information between microstate and trajectory type approaches the equilibrium system entropy:
\begin{equation}
\lim_{\Nmeso \to N_{\rm micro}} I[\St;\Subens] = H(\St) \>.
\end{equation}

\subsection{Construction of one-dimensional reaction coordinates}

In two-mesostate transition-path theory, the (scalar) committor plays an important role as the one-dimensional reaction coordinate that is fully informative about the reaction. The committor can be used to coarse-grain the configuration space into isocommittor surfaces~\cite{E2006}, and determining properties of the system (e.g., mean values of particular system coordinates that are relevant to the reaction) on the isocommittor surfaces yields insight into the reaction mechanism. A particularly interesting surface is the $q=0.5$ isocommittor, the transition-state ensemble of the reaction. 

Consider how this picture changes in multi-mesostate transition-path theory~\cite{Schutte2011}. The $i \to j$ and $j \to i$ transition-path ensembles are characterized by two committor components $q_i(\st)=p(\Org=\meso_i|\st)=p(\Out=\meso_i|\st)$ and $q_j(\st)=p(\Org=\meso_j)=p(\Out=\meso_j|\st)$. In two-mesostate transition-path theory (where without loss of generality $i=A$ and $j=B$), these components are related by $q_A(\st)+q_B(\st)=1$ and there is an obvious choice of scalar reaction coordinate; such a relationship does not exist with multiple mesostates where the $i$ and $j$ committor components do not sum to unity ($q_i(\st)+q_j(\st) \le 1$) and their sum may change throughout configuration space ($q_i(\st)+q_j(\st) \neq q_i(\st') + q_j(\st')$ for all $\st$ and $\st'$). Nevertheless, a local reaction coordinate $\RC_{i j}(\st)$ can be constructed as a function of both committor components
\begin{align}
\label{eq:pairwise_committor}
    \RC_{i \to j}(\st) &= \frac{q_j(\st)}{q_i(\st)+q_j(\st)} \>.
\end{align}
which we term the ``pairwise committor``~\cite{Tan2018} for the $i \to j$ reactive subensemble. In line with the two-mesostate committor function, the pairwise committor has values in $[0,1]$, where microstates with pairwise committor closer to $0$ are dynamically closer to $\meso_i$ (``reactant-like'') and microstates with pairwise committor closer to $1$ are dynamically closer to $\meso_j$ (``product-like''). At equilibrium, the reaction coordinate for the reverse transition (with reactant $\meso_j$ and product $\meso_i$) is $\RC_{j \to i}=1-\RC_{i \to j}$. Analogous to the two-mesostate transition-state ensemble, the $i \to j$ transition-state ensemble is identified as microstates with $\RC_{i \to j}(\st)=0.5$, where the system has equal probability to next reach $\meso_i$ or $\meso_j$. Other isosurfaces of the pairwise committor span microstates that have the same odds of next reaching $\meso_j$ before $\meso_i$. One can construct ${\Nmeso \choose 2}$ such local reaction coordinates to describe each of the reactive subensemble pairs. These reaction coordinates are not independent, since only $\Nmeso-1$ committor functions are needed to construct them.

\section{System-subensemble dynamics}

\subsection{Microstate dynamics in trajectory-outcome and trajectory-origin subensembles}

We consider the dynamics in the joint space of microstates and trajectory types, which yields the stationary joint distribution~\eqref{eq:joint_distr_st_org_out}. The joint transition rates for the categorical outcome and origin and the system microstate are~\cite{Vanden-Eijnden2014,Louwerse2022}
\begin{equation} 
	\label{eq:TPE_transition_rates}
	T^{\subens \subens'}_{\st \st'} =
	\begin{cases} 
	T^{\subens}_{\st \st'} \equiv T_{\st \st'} \frac{p(\out| \st)}{p(\out| \st')}, & \subens' = \subens \\
	T_{\st \st'} p(\Out=\meso_j|\st) \, , &
	\begin{cases}
	\st' \in \meso_i \> {\rm{, }} \> \st \notin \meso_i \> \rm{, } \\
	\subens'=(\meso_i,\meso_i) \> {\rm{, }} \\
        \subens=(\meso_i,\meso_j) \\
	\end{cases} \\
	\vspace{-2.5ex}\\
	T_{\st \st'}/p(\Out=\meso_i|\st') \, , &
	\begin{cases}
	\st' \notin \meso_i \> {\rm{, }} \> \st \in \meso_i \> \rm{, } \\
	\subens'=(\meso_j,\meso_i) \> {\rm{, }} \\
        \subens=(\meso_i,\meso_i)  \\
	\end{cases} \\
	\vspace{-2.5ex}\\
	-\sum
	\limits_{\substack{\st'' \neq \st'\\ \subens'' \neq \subens'}} 
	\, T_{\st'' \st'}^{\subens'' \subens'}, & \st = \st' \> {\rm{,}} \> \subens = \subens' \\
	0 & \rm{otherwise} 
	\end{cases}
	\!\!.
\end{equation}
The top transition rate is for transitions that do not change the trajectory type $\subens$, the second is for transitions that leave a mesostate $\meso_i$ to start a reactive trajectory to $\meso_j$, the third is for transitions that finish a reactive trajectory from $\meso_j$ to $\meso_i$, and the fourth ensures that at steady state the probability is conserved for each microstate.

\subsection{Information flows}

As the system evolves at equilibrium, the respective changes in each of the random variables contribute to changes in the mutual information~\eqref{eq:mutual_information} between system microstate and trajectory type, such that they on average cancel to result in no change to the mutual information. We consider separately the changes in mutual information during changes of trajectory type, $\dot{I}^{\Subens}[\St;\Subens]$, and during microstate transitions within a fixed trajectory type, $\dot{I}^{\St}[\St;\Subens]$. These two contributions are equal in magnitude and have opposite sign due to stationarity of total mutual information at equilibrium,
\begin{equation}
0 = \md_t I[\St;\Subens] = \dot{I}^{\Subens}[\St;\Subens] + \dot{I}^{\St}[\St;\Subens] \>.
\end{equation}
We now show now that the two contributions in fact vanish separately, due to time-reversal symmetry.

The rate of change in mutual information during microstate transitions that do not change the trajectory type is
\begin{subequations}
\begin{align}
\dot{I}^{\St}[\St;\Subens] &= \sum_{\org} \sum_{\out} \sum_{\st,\st'} J^{\subens}_{\st \st'} \ln \frac{p(\out|\st) p(\org|\st)}{p(\out|\st') p(\org|\st')} \\
&= \sum_{\org} \sum_{\st,\st'} T_{\st \st'} \pi(\st') p(\org|\st') \ln \frac{p(\org|\st)}{p(\org|\st')} \nonumber \\
&\quad + \sum_{\out} \sum_{\st,\st'} T_{\st \st'} \pi(\st') p(\out|\st) \ln \frac{p(\out|\st)}{p(\out|\st')} \\
&= \dot{I}^{\St}[\St;\Org] + \dot{I}^{\St}[\St;\Out] \\
&= 0 \ ,
\end{align}
\end{subequations}
which we split into information generated about the outcome, $\dot{I}^{\St}[\St;\Out] \ge 0$, and about the origin, $\dot{I}^{\St}[\St;\Org] \le 0$, that are equal in magnitude and have opposite sign due to time-reversal symmetry of trajectory outcome and origin at equilibrium. The rate of change in mutual information during changes of trajectory type is
\begin{subequations}
\begin{align}
\dot{I}^{\Subens}&[\St;\Subens] = \sum_i \sum_{j \neq i} \sum_{\st \notin \meso_i} \sum_{\st' \in \meso_i} \Bigg[ T_{\st \st'} \pi(\st') p(\Out=\meso_j|\st) \nonumber \\
&\quad \quad \times \ln \frac{p(\Out=\meso_j|\st) p(\Org=\meso_i|\st) p(\meso_i,\meso_i)}{p(\meso_i,\meso_j)} \nonumber \\
&\quad + T_{\st' \st} \pi(\st) p(\Org=\meso_j|\st) \\
&\quad \quad \times \ln \frac{p(\meso_j,\meso_i)}{p(\Out=\meso_i|\st) p(\Org=\meso_j|\st) p(\meso_i,\meso_i)} \Bigg] \nonumber \\
&= 0 \ ,
\end{align}
\end{subequations}
where the first term is the contribution from transitions where the outcome changes from $\meso_i$ to $\meso_j$ (system starts a reactive trajectory from $\meso_i$ to $\meso_j$) and the second term is the contribution from transitions where the origin changes from $\meso_j$ to $\meso_i$ (system finishes a reactive trajectory from $\meso_j$ to $\meso_i$). These terms are equal in magnitude and have opposite sign since the two sets of transitions leading to changes of trajectory type are time-reversed counterparts.

\subsection{Entropy changes}

The microstate transitions that do not change trajectory type $\subens$ are in general time-reversal asymmetric, which can be quantified by a local detailed-balance relation,
\begin{equation}
    \frac{T^{\subens}_{\st \st'} p(\st',\subens)}{T^{\subens}_{\st' \st} p(\st,\subens)} = \frac{p(\Out=\meso_j|\st)p(\Org=\meso_i|\st')}{p(\Out=\meso_j|\st')p(\Org=\meso_i|\st)} \>.
\end{equation}
For stationary subensembles, $p(\Out=m_i|\st)=p(\Org=m_i|\st)$ and the ratio equals unity, so there is no time-reversal asymmetry. For reactive subensemble pairs, the ratio can be expressed as a change in the reaction coordinate for that subensemble,
\begin{subequations}
\begin{align} 
    \ln &\frac{p(\Out=\meso_j|\st)p(\Org=\meso_i|\st')}{p(\Out=\meso_j|\st')p(\Org=\meso_i|\st)} \nonumber \\
    &\quad \quad = \ln \frac{\RC_{i \to j}(\st)}{\RC_{j \to i}(\st)} - \ln \frac{\RC_{i \to j}(\st')}{\RC_{j \to i}(\st')} \\
    &\quad \quad = \ln \frac{\RC_{i \to j}(\st)}{1-\RC_{i \to j}(\st)} - \ln \frac{\RC_{i \to j}(\st')}{1-\RC_{i \to j}(\st')} \>.
\end{align}
\end{subequations}
Thus we can consider time-asymmetric reactive dynamics as arising from a nonconservative potential $\ln \RC_{i \to j}(\st')/(1-\RC_{i \to j}(\st'))$ on the system, that is a function of the reaction coordinate. 

To understand the effect of this nonconservative potential on the system thermodynamics, consider the joint entropy $H(\St,\Subens)\equiv-\sum_{\st,\subens} p(\st,\subens) \ln p(\st,\subens)$. The change in joint entropy can be decomposed into three terms~\cite{Busiello2020}:
\begin{subequations}
\begin{align}
    0 &= \md_t H(\St,\Subens) \\
    &= \underbrace{\sum_{\st,\st', \subens} T^{\subens}_{\st \st'} p(\st',\subens) \ln \frac{T^{\subens}_{\st \st'} p(\st',\subens)}{T^{\subens}_{\st' \st} p(\st,\subens)} }_{\dot{H}^{\rm{irr}}(\St,\Subens)} \nonumber \\
    &\quad - \underbrace{\sum_{\st,\st', \subens} T^{\subens}_{\st \st'} p(\st',\subens) \ln \frac{T^{\subens}_{\st \st'}}{T^{\subens}_{\st' \st}}}_{\dot{H}^{\rm{env}}(\St,\Subens)} \nonumber\\
    &\quad + \underbrace{\sum_{\st,\st', \subens\neq\subens'} T^{\subens \subens'}_{\st \st'} p(\st',\subens') \ln \frac{p(\st',\subens')}{p(\st,\subens)}}_{\dot{H}^{\rm{sub}}(\St,\Subens)}\>, \label{eq:decompose_entropy}
\end{align}
\end{subequations}
where $\dot{H}^{\rm{irr}}(\St,\Subens)$ is the irreversible entropy production, $\dot{H}^{\rm{env}}(\St,\Subens)$ is the environmental entropy change for transitions that do not change the trajectory type, and $\dot{H}^{\rm{sub}}(\St,\Subens)$ is the change in joint entropy due to transitions that change the trajectory type.

Substituting the transition rates that change trajectory type~\eqref{eq:TPE_transition_rates}, the change in joint entropy due to trajectory-type changes is
\begin{subequations} \label{eq:subens_entropy}
\begin{align}
    \dot{H}^{\rm{sub}}&(\St,\Subens) = \sum_{\st,\st',\subens \neq \subens'} T^{\subens \subens'}_{\st \st'} p(\st',\subens') \ln \frac{p(\st',\subens')}{p(\st,\subens)} \\
    &= \sum_{i} \sum_{j \neq i} \sum_{\st \notin \meso_i} \sum_{\st' \in \meso_i} \label{eq:subens_transition} \\
    &\quad \Bigg[ T_{\st \st'} \pi(\st') p(\Out=\meso_j|\st) \ln \frac{p(\st',(\meso_i,\meso_i))}{p(\st,(\meso_i,\meso_j))} \nonumber \\
    &\quad + T_{\st' \st} \pi(\st) p(\Org=\meso_j|\st) \ln \frac{p(\st,(\meso_j,\meso_i))}{p(\st',(\meso_i,\meso_i))} \Bigg]  \nonumber \\
    &= 0 \>.
\end{align}
\end{subequations}
In Eq.~\eqref{eq:subens_transition}, we write the change in joint entropy for all transitions that leave mesostate $\meso_i$ and begin a transition path to $\meso_j$ and for all transitions that enter mesostate $\meso_i$ after completing a transition path from $\meso_j$. Due to time-reversal symmetry, $p(\Out=\meso_j|\st)=p(\Org=\meso_j|\st)$ and $p(\st,(\meso_i,\meso_j))=p(\st,(\meso_j,\meso_i))$ and thus the two entropy changes within square brackets cancel. 

Substituting the joint transition rates~\eqref{eq:TPE_transition_rates} into the expression for the environmental entropy change gives
\begin{subequations} \label{eq:mn_env_entropy}
\begin{align}
    \dot{H}^{\rm{env}}(\St,\Subens) &= \sum_{\st,\st', \subens} T^{\subens}_{\st \st'} p(\st',\subens) \nonumber \\
    &\quad \quad \quad \times \left[ \ln \frac{T_{\st \st'} }{T_{\st' \st}}  + 2 \ln \frac{p(\out|\st)}{p(\out|\st')} \right] \label{eq:mn_split_cond_rate} \\
    &= -\sum_{\subens} p(\subens) \dot{Q}^{\subens} + 2 \dot{I}^{\St}[\St;\Out] \\
    &= 2 \dot{I}^{\St}[\St;\Out] \label{eq:mn_cancel_heat} \>,
\end{align}
\end{subequations}
where $\dot{I}^{\St}[\St; \Out] \ge 0$ is the rate of change in mutual information between the trajectory outcome and system microstate due to system dynamics in a fixed trajectory type, and $\dot{Q}^{\subens}$ is the mean rate of change of system energy (see Eq.~\eqref{eq:Boltzmann_distr}) during a fixed trajectory type $\subens$. Eq.~\eqref{eq:mn_cancel_heat} follows from summing over $\subens$ to cancel average energy changes during forward and reverse transition-path ensembles that are equal in magnitude and have opposite sign. 

The irreversible entropy production~\cite{Seifert_2012} of all subensemble dynamics is
\begin{subequations} \label{eq:epr_is_info}
\begin{align}
&\dot{H}^{\rm{irr}}(\St,\Org,\Out) = \sum_{\subens} p(\subens) \dot{\Sigma}_{\subens} \\
&= \sum_{\org,\out} \sum_{\st,\st'} T^{\subens}_{\st \st'} p(\st',\subens) \nonumber \\
&\quad \times \left[ \ln \frac{T_{\st \st'} \pi(\st')}{T_{\st' \st} \pi(\st)} + \ln \frac{p(\org|\st')}{p(\org|\st)} + \ln \frac{p(\out|\st)}{p(\out|\st')} \right] \\
&= \sum_{\out} \sum_{\st \st'} T_{\st \st'} \pi(\st') p(\out|\st) \ln \frac{p(\out|\st)}{p(\out|\st')} \\
&\quad + \sum_{\org} \sum_{\st \st'} T_{\st \st'} \pi(\st') p(\org|\st') \ln \frac{p(\org|\st')}{p(\org|\st)} \\
&= \dot{I}^{\St}[\St;\Out]-\dot{I}^{\St}[\St;\Org] \>,
\end{align}
\end{subequations}
which equals the rate of information generation about the trajectory outcome, $\dot{I}^{\St}[\St;\Out] \ge 0$, minus the rate of information loss about the trajectory origin, $\dot{I}^{\St}[\St;\Org] \le 0$.

Separately analyzing each individual trajectory subensemble, the entropy production rate within a stationary $\meso_i \to \meso_i$ subensemble is zero, while the entropy production rate within a reactive $\meso_i \to \meso_{j \neq i}$ subensemble is
\begin{subequations}
\begin{align} \label{eq:epr_for_subens}
p(\Org =&\meso_i, \Out=\meso_j) \dot{\Sigma}_{i \to j} \nonumber \\ &= \sum_{\st \st'} T_{\st \st'} \pi(\st') p(\Out=\meso_j|\st) p(\Org=\meso_i|\st') \nonumber \\
&\quad \times \ln \frac{p(\Out=\meso_j|\st) p(\Org=\meso_i|\st')}{p(\Out=\meso_j|\st') p(\Org=\meso_i|\st)} \\
&= \sum_{\st \st'} T_{\st \st'} \pi(\st') p(\Out=\meso_j|\st) p(\Org=\meso_i|\st') \nonumber \\
&\quad \times [ \ln \frac{\RC_{i \to j}(\st)}{1-\RC_{i \to j}(\st)}-\ln \frac{\RC_{i \to j}(\st')}{1-\RC_{i \to j}(\st')} ] \\ 
&\ge 0 \>,
\end{align}
\end{subequations}
which equals the flux-weighted average of changes to the local reaction coordinate. The forward and reverse transition-path ensembles for a given pair of mesostates have equal 
entropy production rates,
\begin{subequations}
\begin{align}
p(\meso_i,\meso_j) \dot{\Sigma}_{i \to j} &= p(\meso_j,\meso_i) \dot{\Sigma}_{j \to i} \\
\dot{\Sigma}_{i \to j} &= \dot{\Sigma}_{j \to i} \>,
\end{align}
\end{subequations}
where we use $p(\meso_i,\meso_j)=p(\meso_j,\meso_i)$.

\section{Multipartite entropy production rates and coordinate relevance to the reaction}

We now consider how the entropy production rates in a given subensemble inform us about system degrees of freedom relevant to the local reaction mechanism. Suppose the system microstate $\St$ is described by $d$ degrees of freedom $\bm{X}=\{ X_1, X_2, \ldots X_d\}$. We further assume that the dynamics are multipartite, such that only one $X_{\nu}$ changes in a given timestep,
\begin{equation} 
T_{\bm{x} \bm{x}'} = 
\begin{cases}
    T_{\bm{x} \bm{x'}}
    &  x_{\nu} \neq x_{\nu}' \> {\rm{,}} \> x_{\mu} = x_{\mu}' \, \forall \mu \neq \nu \\
    -\sum \limits_{\substack{\bm{x}'' \neq \bm{x}'\\}} 
    \, T_{\bm{x}'' \bm{x}'} & \bm{x} = \bm{x}' \\
    0 & \rm{otherwise}
\end{cases} \>.
\end{equation}

The irreversible entropy production of all reactive subensemble pairs can be split into contributions from each degree of freedom $X_{\nu}$,
\begin{align}
\sum_{\out} \sum_{\org \neq \out} &p(\org,\out) \dot{\Sigma}_{\org,\out} \nonumber \\
&= \sum_{\out} \sum_{\org \neq \out} \sum_{{\nu}} p(\org,\out) \dot{\Sigma}^{X_{\nu}}_{\org,\out} \ ,
\end{align}
and the entropy production rate of coordinate $X_{\nu}$ equals the information generation rate due to that coordinate's dynamics,
\begin{subequations}
\begin{align}
\sum_{\out} \sum_{\org \neq \out} p(\org,\out) &\dot{\Sigma}^{X_{\nu}}_{\org,\out} = 2 \dot{I}^{X_{\nu}} [\St;\Out] \\
&= \dot{I}^{X_{\nu}} [\St;\Out] - \dot{I}^{X_{\nu}} [\St;\Org] \>,
\end{align}
\end{subequations}
following analogous steps as in Eq.~\eqref{eq:epr_is_info}.

The entropy production rate due to $X_{\nu}$ dynamics quantifies the contribution of this coordinate to the irreversibility of all reactive subensemble pairs,
\begin{subequations}
\begin{align}
\dot{\Sigma}^{X_{\nu}}_{\org,\out} &= \sum_{\bm{x}, \bm{x}'} \frac{T_{\bm{x} \bm{x'}} \pi(\bm{x}') p(\out|\bm{x}) p(\org|\bm{x}') }{p(\org,\out)} \nonumber \\
&\quad \quad \times \ln \frac{p(\out|\bm{x}) p(\org|\bm{x}')}{p(\out|\bm{x}') p(\org|\bm{x})} \>,
\end{align}
\end{subequations}
where the summation runs over all $\bm{x}$ and $\bm{x'}$ such that $x_{\nu} \neq x_{\nu}' \> {\rm{,}} \> x_{\mu} = x_{\mu}' \, \forall \mu \neq \nu$.

One can find coordinates that are relevant to characterizing all reactive subensemble pairs in the network, or focus on a specific transition and identify a coordinate with maximal entropy production in that particular transition-path ensemble of interest. A one-dimensional coordinate with maximal entropy production rate in the $i \to j$ transition-path ensemble is the local reaction coordinate (or any invertible transformation thereof, e.g., the pairwise committor Eq.~\eqref{eq:pairwise_committor}), such that
\begin{equation}
\dot{\Sigma}_{\org,\out} = \dot{\Sigma}^{\RC_{i \to j}}_{\org,\out} \>.
\end{equation}
The irreversible entropy production thus provides a meaningful quantitative criterion that determines the informative value of any system coordinate to describing a particular transition.

\subsection{Information geometry} \label{sec_info_geometry}

The committor vector suggests an $\Nmeso$-dimensional ``reaction space'', where each axis corresponds to the conditional probability $p(\Out=\meso_i|\st)$ (or equivalently $p(\Org=\meso_i|\st)$) for each mesostate $\meso_i$, ranging from $0$ to $1$. The committor vector of a microstate can be interpreted as a point in the reaction space, with mesostates making up the vertices ($p(\Org=\meso_i|\st \in \meso_i)=1$). Since the conditional probabilities of each origin must sum to unity, the accessible reaction space is the $(\Nmeso\!\!-\!\!1)$-dimensional probability simplex. Here, we show that an information metric exists in this space that is related to the average entropy production rate of all transition-path ensembles.

In Appendix~\ref{appendix_info_geo}, we show that the entropy production rate~\eqref{eq:epr_is_info} for each forward and reverse transition-path ensemble pair can be re-written as a contribution to a Kullback-Leibler divergence $D_{\rm KL}$ for each transition~\cite{Cover2006}:
\begin{subequations}
\begin{align}
&\sum_{\subens} p(\subens) \dot{\Sigma}_{\subens} = \dot{I}^{\St}[\St;\Out]-\dot{I}^{\St}[\St;\Org] \\ 
&= \sum_{\st \st'} T_{\st \st'} \pi(\st') D_{\rm KL} [p(\subens|\st') || p(\subens|\st)] \\
&\approx \frac{1}{2} \sum_{\st \st'} T_{\st \st'} \pi(\st') [x_{\nu}-x'_{\nu}] \mathcal{I}_{\mu \nu} (\st') [x_{\mu}-x'_{\mu}] \label{eq:state_FI} \ ,
\end{align}
\end{subequations}
where $x_{\nu}$ is the $\nu$th component of the vector $\bm{X}$ parameterizing the multidimensional system microstate, and 
\begin{subequations} \label{eq:Fisher_info}
\begin{align}
\mathcal{I}_{\mu \nu}(\st) &\equiv \sum_{\subens} p(\subens|\st) \frac{\partial \ln p(\subens|\st)}{\partial x_{\mu}} \frac{\partial \ln p(\subens|\st)}{\partial x_{\nu}} \\
&= \sum_i \frac{2}{q_i(\st)} \frac{\partial q_i(\st)}{\partial x_{\mu}} \frac{\partial q_j(\st)}{\partial x_{\nu}}
\end{align}
\end{subequations}
is the Fisher information metric~\cite{Cover2006} at microstate $\st$, quantifying the curvature of the probability simplex at that point.

The Fisher information metric defines a square distance
\begin{equation}
\md l^2_{\st \st'} \equiv \frac{1}{2} [x_{\mu}-x'_{\mu}] \mathcal{I}_{\mu \nu} (\st') [x_{\nu}-x'_{\nu}] 
\end{equation}
in the reaction space for a transition from $\st' \to \st$. Substituting this square distance into Eq.~\eqref{eq:state_FI} gives
\begin{align}
    \sum_{\subens} p(\subens) \dot{\Sigma}_{\subens} &\approx \sum_{\st \st'} T_{\st \st'} \pi(\st') \md l^2_{\st \st'} \>. 
\end{align}
The distance $\md l_{\st \st'}^2$ between two microstates is the mean transition-path entropy-production rate averaged over all subensembles, a measure of the contribution of that microstate transition to the irreversibility of the joint dynamics of system and trajectory type.

\section{System dynamics within trajectory-origin subensembles}

Throughout this work, we have focused on trajectory subensembles defined by both trajectory outcome and origin, which highlight reactive trajectory segments in the underlying dynamics. However, a more typical dynamical coarse-graining specifies only the trajectory origin~\cite{Berezhkovskii2019}. This type of coarse-graining arises in milestoning, a method originally developed to sample the transition region between two metastable states by choosing mesostates (``milestones") at discrete intervals on a one-dimensional reaction coordinate~\cite{Elber2004}, or as edges of a grid in multidimensional spaces~\cite{Elber2020}. By initiating trajectories at one mesostate and collecting statistics on the time to reach the next mesostate and the identity of the next mesostate, one builds a coarse-grained model of the dynamics in the transition regions which yields the overall rate constant for a two-mesostate reaction. Further, milestoning can be used to build a coarse-grained model of dynamics between mesostates representing metastable states of the system~\cite{Schutte2011,Buchete2008}, similar to those expressed in Markov-state models~\cite{Chodera2014,Pande2010,Prinz2011}.

Importantly, ``states'' in a coarse-grained model built from milestoning are defined by the trajectory origin~\cite{Berezhkovskii2019} and not the current system microstate. Therefore the marginal dynamics in the space of the trajectory origin constitutes a coarse-graining of \textit{trajectories} and not a coarse-graining of \textit{microstates}. This contrasts with Markov-state models where the configuration space is coarse-grained to fully partition it into mesostates~\cite{Chodera2014,Hartich2020}, and coarse-grained dynamics are constructed by observing the frequency of transitions across the boundaries between mesostates. 

We adapt the framework presented here to trajectory-origin dynamics. Partitioning the long equilibrium trajectory into segments with the same trajectory origin yields $\Nmeso$ subensembles, each defined by a particular trajectory origin. Since the underlying dynamics are Markovian and do not depend on trajectory origin, the joint transition rate for a $(\st',\org') \to (\st,\org)$ transition is $T_{\st \st'}$, unmodified from the original dynamics, in contrast to Eq.~\eqref{eq:TPE_transition_rates} which also specifies the trajectory outcome. The subensembles with common trajectory origin have a net flux between microstates, since trajectories that were at mesostate $\meso_i$ will eventually reach another mesostate. In Appendix~\ref{appendix_traj_origin}, we expand the change in the joint entropy $H(\st,\org)$ to quantify the irreversible entropy production,
\begin{subequations}
\begin{align}
&\dot{H}^{\rm irr}(\St,\Org) = \sum_{\st, \st'} \sum_{\org} T_{\st \st'} \pi(\st') p(\org|\st') \ln \frac{T_{\st \st'} p(\st',\org)}{T_{\st' \st} p(\st,\org)} \\
&= \sum_{i} \sum_{\st, \st' \notin \meso_{j \neq i}} \!\!\!\! T_{\st \st'} \pi(\st') p(\Org=\meso_i|\st') \ln \frac{p(\Org=\meso_i|\st')}{p(\Org=\meso_i|\st)} \\
&= -\dot{I}^{\St}[\St;\Org]  \label{eq:origin_irr_epr} \\
&\ge 0 \>, 
\end{align}
\end{subequations}
where $\dot{I}^{\St}[\St;\Org]$ is the change (here a decrease) in information between system state and trajectory origin due to system transitions within the same trajectory-origin subensemble. Thus the equivalence between irreversible entropy production and information flows in trajectory subensembles is maintained when only the trajectory origin is used to partition the system's dynamics. A similar equality can be found for trajectory-outcome dynamics. 

The entropy production rate within specific subensemble $\Org=\meso_i$ is
\begin{align}
p(\Org=\meso_i) \dot{\Sigma}_{\meso_i} = &\sum_{j \neq i} \sum_{\st \notin \meso_j} \sum_{\st'} T_{\st \st'} \pi(\st') p(\Org=\meso_i|\st') \nonumber \\
&\times \ln \frac{p(\Org=\meso_i|\st')}{p(\Org=\meso_i|\st)} \>.
\end{align}
The origin-subensemble entropy production rate can be interpreted as indicating the irreversibility of dynamics for the system leaving mesostate $\meso_i$, and can be used to find coordinates whose dynamics are most responsible for loss of information about the trajectory origin. Coarse-graining by partitioning a long trajectory rather than partitioning configuration space therefore allows identification of underlying microscopic dynamics that are informative about the coarse-grained transitions, providing a deeper understanding that spans both mesoscopic and microscopic scales.

\section{Discussion}

For reaction networks with many distinct conformations, determining the relative stability of key species (thermodynamics) and the rate of transitions between them (kinetics) allows a global understanding of the function of the system. In this paper, we investigated how the underlying microscopic dynamics provide information about the reactions (mechanisms) in the system by combining tools from information theory and stochastic thermodynamics with multi-mesostate transition-path theory.

The main significance of the results is a quantitative criterion for reaction-coordinate identification. In two-mesostate transition-path theory, it is well known that the committor is the ``perfect'' reaction coordinate, which we clarify as a sufficient statistic in an information-theoretic sense. In a multi-mesostate transition-path network, we find that the statements are slightly modified. An $\Nmeso$-dimensional committor vector is fully informative about the entire reaction network, and committor components can be combined to yield ${\Nmeso \choose 2}$ reaction coordinates, one for each conjugate pair of mesostate transitions. Each is a `local' reaction coordinate that provides a distinctive measure of the system components relevant for a particular transition-path ensemble.

The transition-path entropy production quantifies the information that system dynamics generate, resolving whether the system is heading to $\meso_i$ or $\meso_j$ next, and the rate at which information about the most recent mesostate is lost. This can help with identifying coordinates that describe one transition, allowing for directed manipulation of a small number of system components that can have an impact on coarse-grained kinetics. While we express this entropy production in terms of committor-vector elements, we stress that determination of the committor is not necessary to calculate entropy production. Entropy production simply measures the imbalance in forward and reverse flux in an ensemble of trajectories; therefore, one can generate an ensemble of transition paths for the reaction of interest and count the number of forward and reverse transitions that occur between discrete regions of configuration space; this information is sufficient to determine the entropy production. To identify a one-dimensional coordinate, the ensemble can be projected onto a single coordinate, and the flux asymmetry on that coordinate can be estimated; making modifications to that coordinate to maximize entropy production is a general procedure to identify the reaction coordinate for that transition-path ensemble. 

In addition to identification of local reaction coordinates, the mutual information between system microstate and trajectory type quantifies the information the coarse-grained model resolves about the underlying microstates. This information depends on the choice of system mesostates and could potentially provide a criterion for optimizing their location in configuration space. Current procedures for selecting mesostates or reduced-coordinate representations typically depend on capturing the longest relaxation modes in the system, which can be quantified by, \emph{e.g.,} the VAMP score~\cite{Mardt2018} or Kemeny constant~\cite{Koskin2022}. Using static mutual information and information-generation rates could provide an alternative perspective on optimizing the choice of mesostates that focuses more on capturing informative aspects of the system entropy rather than capturing the longest timescales.

The multi-mesostate transition-path description represents a coarse-graining of trajectories rather than of microstates. This builds time-asymmetry into the system dynamics on a fundamental level. We considered here systems at equilibrium, but the results could be generalized to coarse-grain trajectories from nonequilibrium steady-state distributions with underlying time-asymmetry. The key distinction from our present results is that the forward and reverse committor vectors would no longer be identical, and some of the simplifications in this manuscript would no longer apply. For example, the forward and reverse transition-path ensembles between two mesostates would no longer be time-reversed twins and would generally have distinct features which can be captured by different forward and reverse reaction coordinates. However, the transition-path ensemble entropy productions from the two ensembles would still allow for identification of these coordinates. Additionally, there is some suggestion that the coarse-grained mesostate dynamics, built from partitioning the supertrajectory, straightforwardly preserve the underlying entropy production of the microscopic dynamics under coarse-graining~\cite{Hartich2020}; this contrasts with dynamics between mesostates that fully partition the configuration space, where hidden entropy production terms arise~\cite{Esposito2012}. This is an interesting future direction of study. 

We derived this information thermodynamics for a discrete microstate system with Master-equation dynamics. In principle, these results can be generalized to continuous configuration spaces and dynamics, which extends their applicability. The information metric in Section~\ref{sec_info_geometry} can be formally introduced in a continuous space, and provides a measure of information distance travelled by a particular path in configuration space. This may be of interest in finding an ``optimal'' path that represents a transition-path ensemble, currently typically represented by minimum-free-energy or maximum-flux transition paths~\cite{Maragliano2006,Zhao2010}.

\vspace{2ex}
\acknowledgments
We thank Jordan Sawchuk and Johan du Buisson (SFU Physics) for helpful feedback on the manuscript. This work was supported by Natural Sciences and Engineering Research Council of Canada (NSERC) Canada Graduate Scholarships Masters and Doctoral (MDL), NSERC Discovery Grant RGPIN-2020-04950 (DAS), and Tier-II Canada Research Chair CRC-2020-00098 (DAS).

\begin{widetext}
\appendix
\section{Information geometry}
\label{appendix_info_geo}

The entropy production rate for each forward and reverse transition-path ensemble pair~\eqref{eq:epr_for_subens} can be re-written as a contribution to a Kullback-Leibler divergence $D_{\rm KL}$ for each transition~\cite{Cover2006}:
\begin{subequations}
\begin{align}
\sum_{\subens} p(\subens) &\dot{\Sigma}_{\subens} = \dot{I}^{\St}[\St;\Out]-\dot{I}^{\St}[\St;\Org] \\
&= \sum_{\st,\st',\out} 
    T_{\st \st'} \pi(\st') p(\out|\st) \ln \frac{p(\out|\st)}{p(\out|\st')} - \sum_{\st,\st',\org} T_{\st' \st} \pi(\st) p(\org|\st) \ln \frac{p(\org|\st')}{p(\org|\st)} 
    \label{eq:write_info_rates} \\
&= \sum_{\st,\st',\out,\org} 
    T_{\st \st'} \pi(\st') p(\out|\st)p(\org|\st) \ln \frac{p(\out|\st)}{p(\out|\st')} - \sum_{\st,\st',\out,\org} T_{\st' \st} \pi(\st) p(\org|\st) p(\out|\st) \ln \frac{p(\org|\st')}{p(\org|\st)} 
    \label{eq:multiply_unity} \\
&= \sum_{\st,\st',\out,\org} 
    T_{\st \st'} \pi(\st') p(\out|\st) p(\org|\st) \ln \frac{p(\out|\st)p(\org|\st)}{p(\out|\st')p(\org|\st')} \label{eq:combine_terms} \\
&= \sum_{\st \st'} T_{\st \st'} \pi(\st') D_{\rm KL} [p(\subens|\st') || p(\subens|\st)] \\
&\approx \frac{1}{2} \sum_{\st \st'} T_{\st \st'} \pi(\st') [x_{\nu}-x'_{\nu}] \mathcal{I}_{\mu \nu} (\st') [x_{\mu}-x'_{\mu}] \ ,
\end{align}
\end{subequations}
where $x_{\nu}$ is the $\nu$th component of the vector $\bm{X}$ parameterizing the multidimensional system microstate, and 
\begin{subequations}
\begin{align}
\mathcal{I}_{\mu \nu}(\st) &\equiv \sum_{\subens} p(\subens|\st) \frac{\partial \ln p(\subens|\st)}{\partial x_{\mu}} \frac{\partial \ln p(\subens|\st)}{\partial x_{\nu}} \\
&= \sum_{ij} q_i(\st) q_j(\st) \frac{\partial \ln q_i(\st) q_j(\st)}{\partial x_{\mu}} \frac{\partial \ln q_i(\st) q_j(\st)}{\partial x_{\nu}} \\
&= \sum_{ij} \frac{1}{q_i(\st) q_j(\st)} \frac{\partial q_i(\st) q_j(\st)}{\partial x_{\mu}} \frac{\partial q_i(\st) q_j(\st)}{\partial x_{\nu}} \\
&= \sum_{ij} \Big[ \frac{q_i(\st)}{q_j(\st)} \frac{\partial  q_j(\st)}{\partial x_{\mu}} \frac{\partial  q_j(\st)}{\partial x_{\nu}} + \frac{\partial q_i(\st) }{\partial x_{\mu}}\frac{\partial  q_j(\st)}{\partial x_{\nu}} + \frac{\partial  q_j(\st)}{\partial x_{\mu}} \frac{\partial q_i(\st) }{\partial x_{\nu}} + \frac{q_j(\st)}{q_i(\st)} \frac{\partial q_i(\st) }{\partial x_{\mu}} \frac{\partial q_i(\st) }{\partial x_{\nu}} \Big] \\
&= \sum_i \frac{2}{q_i(\st)} \frac{\partial q_i(\st)}{\partial x_{\mu}} \frac{\partial q_j(\st)}{\partial x_{\nu}}
\end{align}
\end{subequations}
is the Fisher information metric~\cite{Cover2006} at microstate $\st$.

\section{Trajectory-origin subensembles} \label{appendix_traj_origin}

The transition rate for a $\st' \to \st$ transition within a trajectory-origin subensemble is unchanged from the original system dynamics $T_{\st \st'}$ since they are Markovian. The local detailed-balance relation that quantifies time asymmetry is
\begin{equation}
\frac{T^{\org}_{\st \st'} p(\st',\org)}{T^{\org}_{\st' \st} p(\st,\org)} = \frac{p(\org|\st')}{p(\org|\st)} \>.
\end{equation}
Therefore a $\st' \to \st$ transition where $p(\org|\st)$ changes has some time asymmetry in the trajectory-origin subensembles. The net flux for a $\st' \to \st$ transition within subensemble $\org$ is
\begin{subequations}
\begin{align}
J^{\org}_{\st \st'} &= T_{\st \st'} \pi(\st') p(\org|\st') - T_{\st \st'} \pi(\st') p(\org|\st') \\
&= T_{\st \st'} \pi(\st') \left[ p(\org|\st')-p(\org|\st) \right] \>.
\end{align}
\end{subequations}
Thus there is net flux within the subensemble toward microstates with lower values of $p(\org|\st)$; that is, a net flux away from the origin mesostate $\Org=\meso_i$ towards the $M-1$ other mesostates, which act as absorbing boundary conditions for dynamics in subensemble $\Org=\meso_i$. To maintain steady state within the subensemble, trajectories are regenerated in mesostate $\meso_i$ with flux equal to the total flux of trajectories departing $\Org=\meso_i$ to all other mesostates,
\begin{equation}
\sum_{j \neq i} \nu_{i \to j} \>.
\end{equation}

Now consider the change in joint entropy $H(\St,\Org)$. As before, the joint entropy is split into three contributions
\begin{subequations}
\begin{align}
0 &= \md_t H(\St,\Org) \\
&= \sum_{\st, \st'} \sum_{\org, \org'} T_{\st \st'} \pi(\st') p(\org'|\st') \ln \frac{p(\st',\org')}{p(\st,\org)} \\
&= \underbrace{\sum_{\st, \st'} \sum_{\org} T_{\st \st'} \pi(\st') p(\org|\st') \ln \frac{T_{\st \st'} p(\st',\org)}{T_{\st' \st} p(\st,\org)}}_{\dot{H}^{\rm irr}} - \underbrace{\sum_{\st, \st'} \sum_{\org} T_{\st \st'} \pi(\st') p(\org|\st') \ln \frac{T_{\st \st'}}{T_{\st' \st}}}_{\dot{H}^{\rm env}} \nonumber \\
&\quad + \underbrace{\sum_{\st, \st'} \sum_{\org \neq \org'} T_{\st \st'} \pi(\st') p(\org'|\st') \ln \frac{p(\st',\org')}{p(\st,\org)}}_{\dot{H}^{\rm sub}} \>.
\end{align}
\end{subequations}

The subensemble transitions contribute
\begin{subequations}
\begin{align}
\dot{H}^{\rm sub}(\Phi,\Org) &= \sum_{\st, \st'} \sum_{\org \neq \org'} T_{\st \st'} \pi(\st') p(\org'|\st') \ln \frac{\pi(\st') p(\org'|\st')}{\pi(\st) p(\org|\st)} \\
&= \sum_{i} \sum_{\st \in \meso_{j \neq i}} \sum_{\st' \notin \meso_{j \neq i}} T_{\st \st'} \pi(\st') p(\Org=\meso_i|\st') \ln \frac{\pi(\st') p(\Org=\meso_i|\st')}{\pi(\st)} \label{eq:write_mi_terms} \\
&= \sum_{i} \sum_{\st \in \meso_{j \neq i}} \sum_{\st' \notin \meso_{j \neq i}} T_{\st \st'} \pi(\st') p(\Org=\meso_i|\st') \ln \frac{\pi(\st')}{\pi(\st)} \nonumber \\
&\quad + \sum_{i} \sum_{\st \in \meso_{j \neq i}} \sum_{\st' \notin \meso_{j \neq i}} T_{\st \st'} \pi(\st') p(\Org=\meso_i|\st')  \ln p(\Org=\meso_i|\st') \label{eq:sep_energy_info} \>.
\end{align}
\end{subequations}
In Eq.~\eqref{eq:write_mi_terms}, we explicitly write the change in entropy for transitions leaving the trajectory origin subensemble for $\meso_i$ by entering mesostate $\meso_{j \neq i}$. In Eq.~\eqref{eq:sep_energy_info}, we identify two contributions, one arising from the change in system entropy (internal energy) during the transition and the other a change in conditional entropy of the current subensemble, which is reduced since a new mesostate is reached in the transition, thereby eliminating uncertainty about the trajectory type.

The environment entropy change for transitions within a subensemble is
\begin{subequations}
\begin{align}
\dot{H}^{\rm env}(\Phi,\Org)
&= \sum_{\st, \st'} \sum_{\org} T_{\st \st'} \pi(\st') p(\org|\st') \ln \frac{T_{\st \st'}}{T_{\st' \st}} \\
&= \sum_{i} \sum_{\st, \st' \notin \meso_{j \neq i}} T_{\st \st'} \pi(\st') p(\Org=\meso_i|\st') \ln \frac{T_{\st \st'}}{T_{\st' \st}} \>,
\end{align}
\end{subequations}
which is the average energy change in the system for dynamics in each trajectory-origin subensemble. This cancels the first term in the expression for subensemble entropy change~\eqref{eq:sep_energy_info}, since these two contributions account for all energy changes at steady state, which by definition is zero.

The irreversible entropy production for transitions within a subensemble is
\begin{subequations}
\begin{align}
\dot{H}^{\rm irr}(\Phi,\Org)
&= \sum_{\st, \st'} \sum_{\org} T_{\st \st'} \pi(\st') p(\org|\st') \ln \frac{T_{\st \st'} p(\st',\org)}{T_{\st' \st} p(\st,\org)} \\
&= \sum_{i} \sum_{\st, \st' \notin \meso_{j \neq i}}  T_{\st \st'} \pi(\st') p(\Org=\meso_i|\st') \ln \frac{p(\Org=\meso_i|\st')}{p(\Org=\meso_i|\st)} \\
&= -\dot{I}^{\St}[\St;\Org] \label{eq:compare_to_full_subens} \\
&\ge 0
\end{align}
\end{subequations}
where $\sum_{\st, \st' \notin \meso_{j \neq i}}$ indicates a sum over microstates $\phi$ and $\phi'$, neither of which is in any mesostate $\meso_j$ other than $\meso_i$, and $\dot{I}^{\St}[\St;\Org]$ is the information change due to system transitions within the same trajectory-origin subensemble.

The irreversible entropy production within a subensemble $\Org=\meso_i$ is
\begin{align}
p(\Org=\meso_i) \dot{\Sigma}_{\meso_i} = \sum_{\st, \st' \notin \meso_{j \neq i}} T_{\st \st'} \pi(\st') p(\Org=\meso_i|\st') \ln \frac{p(\Org=\meso_i|\st')}{p(\Org=\meso_i|\st)} \>.
\end{align}
\end{widetext}

%\bibliography{references}

\end{document}